\documentclass[pre,aps,twocolumn]{revtex4}
\usepackage{amsmath,bm,epsfig,graphicx}
 \newcommand{\B}[1]{{\bm{#1}}}

\usepackage[latin1]{inputenc}

\def\BE{\begin{equation}}\def\EE{\end{equation}}
\def\BEA{\begin{eqnarray}}\def\EEA{\end{eqnarray}}
\def\BSE{\begin{subequations}}\def\ESE{\end{subequations}}
\def\<{\left\langle} \def\>{\right\rangle} \def\({\left(} \def\){\right)}
\let \= \equiv

  \let\~\widetilde \let\^\widehat 
   
  \def\1{\bm1}

\def\Fbox#1{\vskip1ex\hbox to 8.5cm{\hfil\fboxsep0.3cm\fbox{%
  \parbox{8.0cm}{#1}}\hfil}\vskip1ex\noindent}  
\begin{document}
\title{Reynolds number dependence of drag reduction by rod-like polymers}

\author{Yacine Amarouch\`ene$^1$, Daniel Bonn$^2$, Hamid Kellay$^1$, Ting-Shek  Lo$^3$,
Victor S. L'vov$^3$ and Itamar Procaccia $^3$}
\affiliation{$^1$CPMOH, U. Bordeaux 1, 351 cours de la Lib\'eration, 33405 Talence, France\\
$^2$ LPS, ENS, Paris, 24 rue Lhomond 75005, Paris France, and van der Waals-Zeeman Instituut, University of Amsterdam, Valckenierstraat 65 1018 XE Amsterdam, The Netherlands\\
$^3$ Dept. of Chemical Physics, The Weizmann Institute of
Science, Rehovot 76100, Israel}
\begin{abstract}
We present experimental and theoretical results addressing the
Reynolds number (Re) dependence of drag reduction by sufficiently
large concentrations of rod-like polymers in turbulent
wall-bounded flows. It is shown that when Re is small the drag is
{\em enhanced}. On the other hand when Re increases the drag is
reduced and eventually the Maximal Drag Reduction (MDR) asymptote
is attained. The theory is shown to be in excellent
 agreement with experiments, rationalizing and explaining all the universal
 and the non-universal aspects of drag reduction by rod-like polymers.
\end{abstract}
\maketitle
\section{Introduction}

The phenomenon of turbulent drag reduction describes the
increase in the throughput  in turbulent flows by adding tiny
amounts of polymers. In this way for instance the amount of liquid
that is transported in a pipe for a given pressure drop can be
increased significantly. This implies a wide range of industrial
applications, and consequently the effect, discovered in 1944 by
Toms \cite{Toms49}, has been studied intensively over the past
decades (see e.g. Ref. \cite{Virk75a}). The aim of this paper is to rationalize and understand the change
from drag enhancement to drag reduction by rod-like polymers in
turbulent wall-bounded flows as a function of the Reynolds number
Re \cite{Book1, ASME, Bonn}. To this aim we present
new experimental data and a theoretical analysis.

\begin{figure}
\includegraphics[width=3.5in]{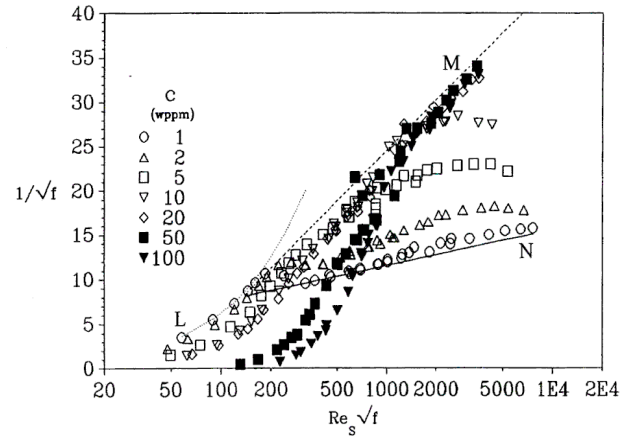}\\
  \caption{The drag in Prandtl-K\'arm\'an coordinates for the water with the rod-like
  polymer PAMH B1120 in 0.0001N NaCl in a pipe flow, see \cite{ASME} for details. The symbols represent the concentrations in wppm (weight parts per million) as given in the figure.  }
  \label{Virk}
\end{figure}

A classical example of the phenomenon
of interest is shown in Fig. \ref{Virk} which pertains to pipe
flow \cite{ASME}. We denote the velocity field $\B U (\B r,t)$ and
its mean over time as $V(y)$ where $y$ is the distance from the
wall. With the mean shear defined as $S(y)\equiv dV(y)/dy$, the
Fanning drag coefficient is defined as
\begin{equation}
f \equiv \tau_w/\Big(\frac{1}{2}\rho \tilde V^2\Big) \ ,
\end{equation}
where $\tau_w$ is the wall shear stress at the wall
\begin{equation}
\tau_w \equiv \rho \nu S(y=0) \ ,
\end{equation}
$\nu$, $\rho$ and $\tilde V$ are the kinematic viscosity, the fluid
density and the mean fluid throughput, respectively. Fig.~\ref{Virk}
presents this quantity as a function of Re in the traditional
Prandtl-Karman coordinates $1/\sqrt{f}$ vs. Re$\sqrt{f}$. The
straight continuous line denoted by `N' presents the Newtonian universal law.
Data points below this line are indicative of a drag enhancement, i.e., an increase in the dissipation due to the addition of the polymer. Converesely, data points above the line correspond to drag reduction, which is
always bound by the Maximal Drag Reduction (MDR) asymptote
represented by the dashed line denoted by `M'. This figure shows data for a
rod-like polymer (a polyelectrolyte in aqueous solution at very low salt concentration) and shows how drag enhancement for low values of Re crosses over to drag reduction at
large values of Re \cite{ASME, Bonn}. One of the results of this paper is the theory presented
below, that reproduces the phenomena shown in Fig. \ref{Virk} in a
satisfactory manner.

It is interesting to note that with {\em flexible} polymers the
situation is very different, and there is no drag enhancement at
any value of Re. The reason for this distinction will be made
clear below as well.

In Sec. 2 we present new experimental results in which drag reduction is measured as a function of the Reynolds number, and for different concentrations of the rod-like
polymer. We determine the rheological properties of these solutions simultaneously, so that the rheology of the solutions is known. In Sec. 3 we review the basic theoretical approach to drag
reduction in general, and by rod-like polymers in particular. We
discuss the effect of the polymers on the fluid flow in three cases:
high Re flows, low Re flows and intermediate Re flows,
respectively. In Sec. 4 we explain how  to solve the
equations and present the results and their comparison to the experiments. Discussions and the
conclusions are presented in Sec. 5.
\section{Experimental}
\subsection{Flow Geometry and Polymer}
In this section we describe the experimental results obtained using well-characterized rod-like
polymer solutions. The turbulence cell has
been previously described in detail by Cadot et al.\cite{Cadot}. A
turbulent flow is generated in a closed cylindrical cell (of volume three liters) between
two counter-rotating disks spaced one disk diameter apart. The disks are
driven by DC motors. The motors are configured to keep the disks rotating at a
constant angular velocity $\Omega$, independently of the torque exerted by the
turbulent fluid on the disks. Thus, all the experiments presented here are
performed at constant angular velocity. All experiments were repeated with pure
water and with the water seeded with the polymers. In both cases we define \cite{Cadot}
the integral Reynolds number as Re$=\rho\Omega R^{2}/\eta_{_{S}}$, where
$\rho$ is the density and $R$ is the radius of the disks. To determine the
energy dissipation in the turbulent fluid (from which we also infer the amount
of drag reduction or drag enhancement), we take advantage of the fact that we are in a closed
system. All the kinetic energy fed into the flow will eventually be dissipated
viscously, leading to a temperature increase of the fluid: by measuring the
temperature increase in time $\Delta T/\Delta t$ we can also estimate the
power dissipated by the turbulent flow. The temperature is measured with a Pt
thermocouple probe with an accuracy of $0.01^\circ$C. From our temperature measurements, we can then deduce the percentage of drag reduction ($DR$) from:
\begin{equation}
DR(\%)\equiv 100\times \left[1-\frac{(\Delta T/\Delta t)_{\rm poly}}{(\Delta T/\Delta t)_{\rm water}}\right] \ .
\end{equation}
For the polymer we use solutions of Xanthane, a stiff rodlike
polymer with an average molecular weight of $M_{w}=3.10^{6}{\rm g.mol}^{-1}$. The
rheology of the polymer solutions was studied on a Reologica Stress-Tech
(cone-plate geometry) rheometer. The latter is equipped with a normal force
transducer, and has a large cone (55 mm) with a small angle ($0.4^\circ$)
in order to be able to detect small normal stress differences at high shear
rates. Changing the concentration allows us to modify the shear-dependent
viscosity of the solutions. The apparent viscosity decreases with
increasing shear rate, i.e., the fluid is shear thinning. A satisfactory
description of most of the data for high shear rates can be obtained using the
Ostwald-de Waele power law model \cite{Bird} for the viscosity $\eta$:
$\eta=k_{1}S^{(n-1)}$; the viscosity shows a power law dependence
on the shear rate $S$. The validity of this model is limited to a
certain range of shear rates, depending on the concentration $C$ of the
polymer solution; the range and constants $k_{1}$ and $n$ are reported in
\cite{Lindner}. For the whole concentration range studied here
($0<C<2000$ wppm), no measurable normal stresses
were found in the range of shear rates studied ($10<S<6000{\rm s}^{-1}$); the uncertainty on the measurements is of the order of 10Pa.

\begin{figure}
\center
  \includegraphics[width=3.5in]{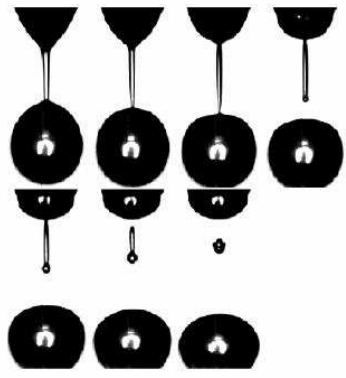}
  \caption{Droplet detachment for a 300 wppm Xanthane solution filmed with a
rapid camera: there is 1ms between subsequent images. }
  \label{droplet}
 \end{figure}

The other important difference in the rheology of the rigid
polymers, compared to solutions of flexible polymers, is the
elongational viscosity $\eta_{E}$, which quantifies the resistance
to stretching of a fluid element. If sufficiently important,
$\eta_{E}$ can be measured \cite{Drops} by looking at droplet
fission with a rapid camera (Fig. \ref{droplet}). The dynamics of
the thinning of the filament that connects the droplet to the
orifice can be used to obtain the elongational viscosity since the
stretching of the filament corresponds to a perfectly elongational
flow. Performing this experiment, we find that, for the rigid
polymer solution, $\eta_{E}$ is so low that the dynamics of the
filament is nearly indistinguishable from that of pure water for
which $\eta_{E}=$3mPas \cite{Trouton}.

\subsection{Results}

\begin{figure}
\center
  \includegraphics[width=3.5in]{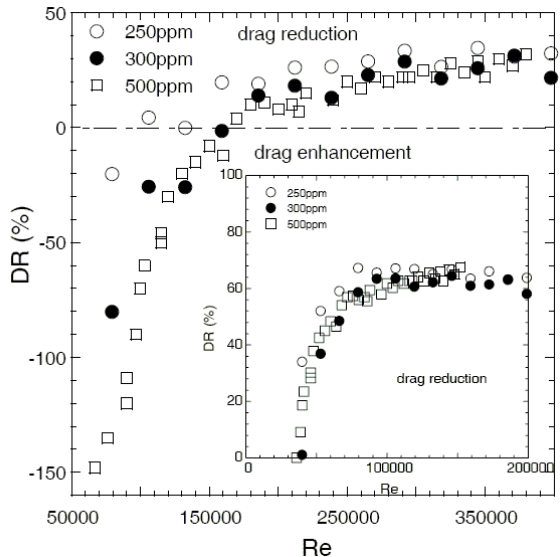}
  \caption{Percentage of drag reduction (or enhancement) vs. Re for three different
  polymer concentrations. Inset: percentage of drag reduction vs. Re when the actual viscosity (including the polymer) is used in the definition of Re). }
  \label{expresults}
 \end{figure}

In Fig. \ref{expresults} the percentage drag reduction $DR(\%)$ is plotted as a
function of the Reynolds number for three different polymer
concentrations. Also in our measurement geometry and with our rod-like polymer at low $Re$ a drag
enhancement is observed, which smoothly
transforms into a drag reduction at high Re.
In this paper we rationalize and explain the results presented in both Fig. \ref{Virk} and Fig. \ref{expresults}. To this aim we turn now to a review of the theory of drag reduction.
\section{Review of the Theory of Drag Reduction by Rod-like polymers}

In the presence of a small concentration of rod-like polymers, the
Navier-Stokes equations for the fluid velocity $\B U(\B r,t)$ gain
an additional contribution to the stress tensor:
\begin{eqnarray}
\label{Navier}
 \frac{\partial \B U}{\partial t} +\B U\cdot \B \nabla \B U&=& - \nabla  p +\nu_0 \nabla^2
\B U  +\nabla \cdot {\B \sigma} \ , \\
\nabla \cdot \B U &=& 0 \ . \nonumber
\end{eqnarray}
The extra stress tensor $\B \sigma$ is due to  the interaction
between the polymers and the fluid. For rod-like polymers it takes
the form \cite{88DE}:
\begin{equation}
\label{stress} \sigma_{ab}=6 \nu_p n_a n_b (n_i n_j
\mathcal{S}_{ij}) \ ,
\end{equation}
where $\nu_p$ is the viscosity contributed by the polymer in the
limit of zero shear, $\B n \equiv \B n(t, \B r)$ is a unit vector
describing the orientation of the polymer, and $S_{ij}\equiv
\partial U_j/ \partial r_j$ is velocity gradient.  The equation of motion for the orientations of the polymers
are approximated by the rigid-dumbbell model which reads
\begin{eqnarray}
\frac{D \mathcal{R}_{ab}}{D t}&=& \mathcal{S}_{ai}
\mathcal{R}_{ib}+\mathcal{S}_{bi}
\mathcal{R}_{ia}-2\mathcal{S}_{ij}
\mathcal{R}_{ij}\mathcal{R}_{ab}
\nonumber \\
&& -6\gamma(\mathcal{R}_{ab}-\frac{\delta_{ab}}{3}) \ ,
\label{R_eqn}
\end{eqnarray}
where $\mathcal{\B R} \equiv \B n \B n$ and $\gamma$ is the
Brownian rotational frequency, proportional to temperature.
\subsection{High Re flows}

The phenomenon of drag reduction  in highly turbulent channel
flows can be understood on the basis of the balance equations of
the mechanical momentum and turbulent energy \cite{PRL}.  For a
channel flow we choose the coordinates  $x$, $y$ and $z$ as the
lengthwise, wall-normal and spanwise directions respectively. The
length and width of the channel are usually taken much larger than
the mid-channel height $L$, making the latter a natural re-scaling
length for the introduction of dimensionless (similarity)
variables, also known as ``wall units" \cite{Pope}. Denoting the
pressure gradient $p'\equiv -\partial p/\partial x $ we define the
friction Reynolds number Re$_\tau$, the normalized distance from
the wall $y^+$ and the normalized mean velocity $V^+(y^+)$ (which
is in the $x$ direction with a dependence on $y$ only) by
\begin{equation}
{\rm Re}_\tau \equiv {L\sqrt{\mathstrut p' L}}/{\nu_0}\ , \  y^+ \equiv
{y {\rm Re}_\tau }/{L} \ , \  V^+ \equiv {V}/{\sqrt{\mathstrut p'L}} \
. \label{red}
\end{equation}
The balance equations are derived on the basis of the Reynolds decomposition
\begin{eqnarray}
&&U_i(\B r,t) = V(y)\delta_{ix} + u_i(\B r, t)\ , \\
&&S_{ij} (\B r,t)= S(y) \delta_{ix}\delta_{jy} +s_{ij}(\B r,t) \ ,
\end{eqnarray}
In addition to the mean shear $S(y)$, we need to to introduce the
mean turbulent kinetic energy $K \equiv \langle u^2 \rangle/2$ and
the Reynolds stress $W \equiv - \langle u_x u_y \rangle$. The
momentum balance equation is obtained by averaging Eq.
(\ref{Navier}) and integrating with respect to $y$, ending up with
the exact equation:
\begin{equation}
\label{raw_mom} \nu_0 S + W + \langle \sigma_{xy} \rangle = p' (L-y) \ .
\end{equation}
The physical meaning of this equation is transparent:  the
momentum generated by the fixed pressure gradient $p'(L-y)$ is
transported to the wall by the momentum flux (=Reynolds stress)
$W$, and dissipated by the Newtonian viscosity $\nu_0 S$.  The
mean stress induced by the polymer ($\langle \sigma_{xy}\rangle $)
is an additional viscous dissipation due to the polymers.

In wall-units Eq. (\ref{raw_mom}) can be written in the form:
\begin{equation}
\label{Meqn} S^+ + W^+ +\langle \sigma^+_{xy} \rangle  = (1-y^+
/{\rm Re}_\tau)\ ,
\end{equation}
where $S^+ \equiv \nu_0 S /(p'L)$,  $W^+ \equiv W/(p'L)$, and
$\sigma_{ij}^+ \equiv \sigma_{ij}/(p'L)$. When $y^+$ not too
large, e.g., in the log-layer,  we neglect the second term on the
RHS, approximating the RHS as unity.

The balance equation for the turbulent kinetic energy is
calculated by taking the dot product of the fluctuation part of
Eq. (\ref{Navier}) with $\B u$:
\begin{equation}
\label{raw_energy} W  S  = \frac{\partial } {\partial y } \langle
u_y u^2  +  u_y p  - \sigma_{iy}u_i  \rangle  + \nu_0 \langle
s_{ij} s_{ij} \rangle \ + \langle \sigma_{ij} s_{ij} \rangle .
\end{equation}
Also this equation is exact. We simplify it by noting that the
first term on the  RHS involves the spatial flux of turbulent
energy which is known to be negligible in the log-layer.  The
second term on the RHS represents the Newtonian dissipation which
is modelled (in wall-units) as \cite{PRL}:
\begin{equation}
\label{K_model} \langle s^+_{ij} s^+_{ij} \rangle \ \approx K^+
\Big(\frac{a}{y^+}\Big)^2 + b\frac{(K^+)^{3/2}}{y^+} \ ,
\end{equation}
where $a$ and $b$ are dimensionless coefficients of the order of
unity and $s_{ij}^+ \equiv \nu_0 s_{ij}/(p'L)$ and $K^+ \equiv
K/(p'L)$. The physical reasons of this model are follows: When
$y^+$ is small, the dissipation of the turbulent energy is
dominated by the viscosity (the firth term).  When $y^+$ is large,
the an eddy with size $y^+$ and energy $K^+(y)$ loses its energy
in the time-scale of $y/\sqrt{K}$, which is the eddy turn-over
time scale. We note that the summation sign in
Eq.(\ref{K_model}) is merely a pseudo sum. It only means that when
$y^+$ is small (large), the first (second) term is more important.
Writing in wall-unit, Eq (\ref{raw_energy}) becomes:
\begin{equation}
\label{Keqn} W^+  S^+  \sim  a^2  \frac{K^+}{(y^+)^2} +  b
\frac{(K^+)^{3/2}}{y^+} + \langle \sigma_{ij}s_{ij} \rangle \ .
\end{equation}

Finally, we quote the experimental fact that in the log-layer
$W^+$ and $K^+$ are proportional to each other:
\begin{equation}
\label{KW} K^+ c^2 = W^+ \ .
\end{equation}
Experimentally, it was found that $c \approx 0.5$ in the Newtonian
case, and $c \approx 0.25$ in the MDR.

To estimate of the polymer term on the LHS of Eq.~(\ref{Meqn}) we
first observe that for large Re when the rod-like polymers are
efficiently aligned by the flow, $R_{xx}\approx 1$. Therefore from
Eq. (\ref{stress}) we conclude that
\begin{equation}
\label{sigma} \sigma_{xy} \sim \nu_p R_{yy} S \ .
\end{equation}
In wall units the momentum equation reads
\begin{equation}
\label{mom} (1 + \nu_p^+ R_{yy}) S^+ + W^+ = 1-\frac{y}{{\rm
Re}_{\tau}}\ ,
\end{equation}
where $\nu_p^+ \equiv \nu_p/\nu_0$.

To estimate the polymer term in the RHS of Eq. (\ref{raw_energy})  (the energy dissipation
due to the polymers), we use the results of \cite{PRE} where it was shown that
\begin{equation}
\langle \sigma_{ij} s_{ij} \rangle \sim \nu_p R_{yy} \frac{K}{y^2} \ .
\end{equation}
Putting into Eq. (\ref{raw_energy}) and writing in wall-unit, we
have
\begin{equation}
\label{energy} W^+  S^+  \sim  a^2 (1 + \nu_p^+ R_{yy})
\frac{K^+}{(y^+)^2} +  b \frac{(K^+)^{3/2}}{y^+} \ .
\end{equation}
Equations  (\ref{mom}) and (\ref{energy}) implies that the
polymers change the properties of the flows by replacing the
viscosity by
\begin{equation}
\label{nu} \nu_{\rm eff} = 1 + \nu_p^+ R_{yy} \ .
\end{equation}
This means that polymeric flows can be reasonably described by the
changing the ``effective viscosity" of the solution. In the fully
developed turbulent flow, it was shown in \cite{PRE} that $R_{yy}$
depends on $K^+$ and $S^+$:
\begin{equation}
\label{turbulent} R_{yy} = \frac{K^+}{(y^+ S^+)^2} \ .
\end{equation}
It was argued in \cite{PRE} that for large Re, $K^+$ grows
linearly with $y^+$ and thus the viscosity profile is linear.
Furthermore, it was shown theoretically in \cite{master} that, if
the effective viscosity varies linearly with $y^+$, i.e.,
\begin{equation}
\label{nu_eff} \nu_{\rm eff}= 1+ \alpha (y^+ - y^+_{\nu}) \ ,
\end{equation}
then $c$ must satisfy the relation
\begin{equation}
\label{master} \frac{a}{c}=\frac{\delta}{1 - \alpha \delta} \ .
\end{equation}
Here $\delta \approx 6$ is the width of the viscous sub-layer in
the Newtonian flows.

For a given $\nu_{\rm eff}$,  the equations (\ref{mom}),
(\ref{energy}), (\ref{KW}) and (\ref{master}) form a complete set
of equations for variables $S^+$, $W^+$, $K^+$ and $c$. This model
has been studied extensively for Newtonian flows \cite{EnFM} and
for the flows with linear viscosity profile \cite{master}. The
resulting velocity profiles agree reasonably with the experimental
results. It was found that in the viscous sub-layer, the model
does allows for a solution that is turbulent, and therefore $K^+=W^+=0$
and $S^+=1$. This also agrees quite well with existing experimental
data. Moreover, an edge solution for $\alpha=1/12$ was observed
\cite{master}.  If $\alpha>1/12$, the model no longer has a solution corresponding to a
turbulent flow and therefore the flow must be laminar. The
special case $\alpha=1/12$ was identified as the MDR of the
polymeric drag reduction.
\subsection{Low Re flows}

According to Eq. (\ref{nu}), the value of $\nu_{\rm eff}$ depends
on $\nu^+_p$ and $R_{yy}$. The value of $\nu^+_p$ is determined by
the polymer properties such as the number of monomers, their concentration etc., and thus $\nu^+_p$
should be considered as an external parameter in the equation. The value of $R_{yy}$,
on the other hand, depends on the properties of the flow. In the
case of laminar flow with a constant shear rate, i.e., $K^+=W^+=0$ and
$S^+= $constant, it was shown theoretically in \cite{88DE} that:
\begin{equation}
\label{laminar} R_{yy} =  \frac{2^{1/3}}{{\rm De}^{2/3}} \ ,
\end{equation}
where the Deborah number De defined by ${\rm De}=S/\gamma$.  Thus,
the effective viscosity is reduced if $S$ is increased, and
therefore the rod-like polymers solution is a shear-thinning
liquid. Naturally, The value of De changes with Re. To clarify this
dependence we consider the momentum equations Eq.(\ref{raw_mom})
at $y=0$ in the Newtonian case.
\begin{equation}
\label{mumbo} \nu_0 S = p'L \ .
\end{equation}
Usually in experiments system size and the
working fluid remain the same.  Therefore, $\nu_0$ and $L$ are
constants and so Re$_\tau$ depends on $p'L$ only. According to
(\ref{red}), Re$_\tau$ grows as $\sqrt{p'L}$ and therefore
\begin{equation}
\label{De} {\rm De} = \frac{\nu_0}{\gamma L^2} {\rm Re}_{\tau}^2
\end{equation}
 by Eq.(\ref{mumbo}).  Putting into Eq.(\ref{laminar}), we have
\begin{equation}
\label{nu_vis} \nu_{\rm eff}= 1+ \nu^+_p \frac{\lambda}{{\rm
Re}_{\tau}^{4/3}} \  ,
\end{equation}
where $\lambda \equiv \nu_0/\gamma L^2 $ is a constant.

\subsection{Intermediate Re flows}

In the case of intermediate Re, we need an interpolation between
Eqs. (\ref{turbulent}) and (\ref{laminar}).  To do this we note
that when $y^+$ is small, the solution of Eqs. (\ref{KW}), (\ref{mom}) and
(\ref{energy}) result in $W^+=K^+=0$ in the viscous
sub-layer. This implies that the flow cannot be highly turbulent
in the viscous sub-layer. Thus, it is reasonable to employ
Eq.(\ref{laminar}) as long as $y^+$ is small.
 On the other hand, as the upper bound of $y^+$
is ${\rm Re}_\tau$, when $y^+$ is large, it automatically implies
that Re$_\tau$ is large. The laminar contribution is therefore
negligible as it varies inversely with Re$_\tau$. The effective
viscosity due to the polymer is dominated by the turbulent
estimate, Eq. (\ref{turbulent}). To connect these two regions we
simply use the pseudo-sum:
\begin{eqnarray}
 \nu_{\rm eff} &=& 1+ \nu^+_p \left(
\frac{\lambda}{{\rm Re}_{\tau}^{4/3}} + \frac{K^+}{(y^+ S^+)^2}
 \right) \nonumber \\ &=& g+ \nu^+_p \frac{K^+}{(y^+ S^+)^2} \label{nu_all} \  ,
\end{eqnarray}
where $g\equiv 1+ \nu^+_p \lambda /{\rm Re}_{\tau}^{4/3}$. One can
see that the limits for both high and low  Re$_\tau$ are satisfied.
\section{Solution and results}
\subsection{Setting up the equations}
Having the expression for $\nu_{\rm eff}$, we have to specify the
value of $c$ in Eq. (\ref{KW}) to complete the equations. This
variable naturally depends on $\alpha$ and $g$.  The
latter dependence, however, can be eliminated by rescaling the
dimensional variables. Define
\begin{equation}
\tilde{{\rm Re}}_\tau \equiv \frac{L\sqrt{\mathstrut p' L}}{g \nu_0}\
, \ \tilde{y}^+ \equiv \frac{y \tilde{{\rm Re}}_\tau }{L} \ , \  V^+
\equiv \frac{V}{\sqrt{\mathstrut p'L}} \ . \label{rescale}
\end{equation}
In these units Eq. (\ref{nu_all}) is written as
\begin{equation}
\label{nu_all2} \tilde{\nu}_{\rm eff} = 1+ \tilde{\nu}_p
\frac{\tilde{K}^+}{(\tilde{y}^+ \tilde{{S}}^+)^2}  \ ,
\end{equation}
where $\tilde{\nu}_p=\nu_p/g$, $\tilde{S}^+=S^+ g$ and
$\tilde{K}^+ = K^+$.  We can find the $\alpha$-dependence by
comparing Eq. (\ref{nu_all2}) to (\ref{nu_eff}) and identifying
the width of the viscous sub-layer $\tilde{y}^+_\nu$ with
$a/c(\alpha)$. This stems from the continuity of $\tilde{S}^+$ at
the boundary of the viscous sub-layer. This means that
\begin{equation}
a/c(\alpha) =\tilde{y}^+_\nu \ . \label{a/c}
\end{equation}
Combining, Eqs (\ref{nu_eff}), (\ref{master}) and (\ref{a/c}), the
relationship between $\tilde{\nu}_{\rm eff}$ and $c$ is:
\begin{equation}
\label{ceqn} \tilde{\nu}_{\rm eff}=1+\frac{a-c
\delta}{a\delta}\Big(\tilde{y}^+-\frac{a}{c}\Big)
\end{equation}
If we note also that Eqs.~(\ref{nu_eff}) and (\ref{nu_all2}) can
be written as
\begin{equation}
\tilde{K}^+ = A^2 (\tilde{S}^+\tilde{y}^+)^2 \label{K}
\end{equation}
with
\begin{equation}
A^2 = \frac{\tilde{\nu}_{\rm eff} -1 }{\tilde {\nu}_p} \ .
\end{equation}

Using then Eqs.~(\ref{KW}) and (\ref{K}), we can rewrite
Eqs.~(\ref{mom}) and (\ref{energy}) as two equations for the two
variables $\tilde{\nu}_{\rm eff}$ and $\tilde{S}^+$:
\begin{equation}
\tilde{\nu}_{\rm eff} \tilde{S}^+ + c^2 A^2 (\tilde{S}^+
\tilde{y}^+)^2 = 1 \ , \label{Neq1}
\end{equation}
and
\begin{equation}
c^2 \tilde{S}^+ = \tilde{\nu}_{\rm eff} (\frac{a}{\tilde{y}^+})^2
+ b A \tilde{S}^+ \ . \label{Neq2}
\end{equation}
Equation (\ref{Neq2}) implies
\begin{equation}
\tilde{S}^+ = \frac{\tilde{\nu}_{\rm eff}}{(\tilde{y}^+)^2} \
\frac{a^2}{(c^2-bA)} \ . \label{S}
\end{equation}
Substituting Eq.~(\ref{S}) into Eq.~(\ref{Neq1}) gives an equation
for $\nu_{\rm eff}$:
\begin{equation}
\tilde{\nu}_{\rm eff}^2 (\frac{a}{\tilde{y}^+})^2 (c^2-bA) +
c^2A^2 \tilde{\nu}_{\rm eff}^2 \Big(\frac{a^2}{\tilde{y}^+}\Big)^2 =
(c^2-bA)^2 \ . \label{Fnueff}
\end{equation}

Finally, we can solve Eqs. (\ref{Fnueff}) and (\ref{ceqn}) to get
$\tilde{\nu}_{\rm eff}(y^+)$ for different values of
$\tilde{\nu}_p$. Then we can obtain $\tilde{S}^+$ and
$\tilde{K}^+$ using Eqs.~(\ref{S}) and (\ref{K}) respectively.
Finally, we reexpress the variables in wall unit by using Eq.
(\ref{rescale}).
\subsection{Comparison of analysis with experiments}

\begin{figure}
  \includegraphics[width=3.5in]{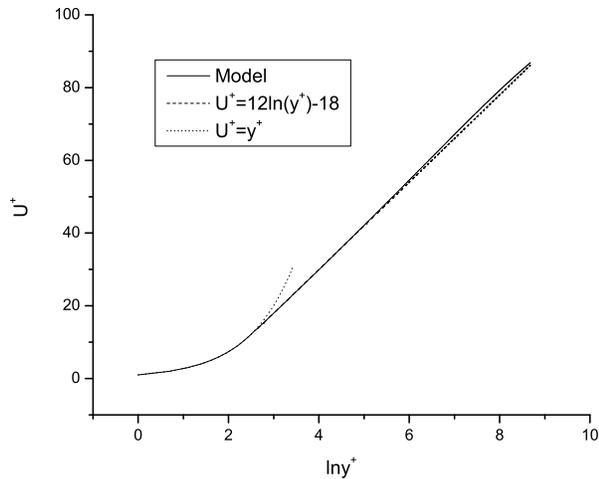}\\
  \caption{The velocity
profile as predicted by the model with $\nu^+_p=10^4$ and
Re$_\tau=590$.}\label{rodMDR}
\end{figure}

To compare our analytical results with the experiments, we first solve the model with
Eq.(\ref{nu_all}) for Re$_\tau$=590, and for $\lambda=0$, i.e. the
high Re limit. This demonstrates that the MDR is reproduced within the model, as is shown in Fig. \ref{rodMDR}.

\begin{figure}[t]
  \includegraphics[width=3.5in]{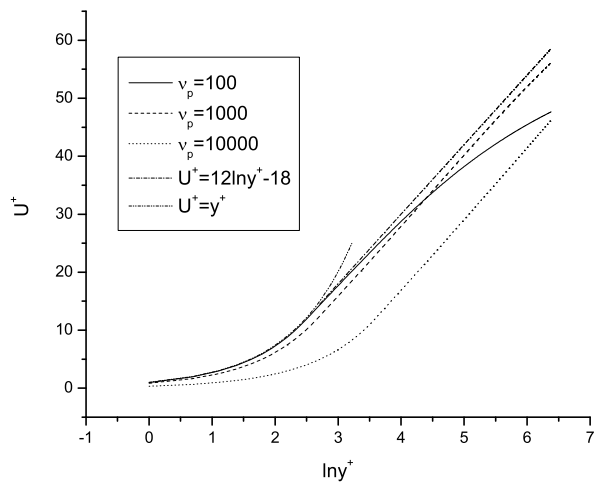}\\
  \caption{The velocity profile predicted by the model with Re$_\tau=590$ and
$\lambda=1$ for various values of $\nu^+_p$.}\label{v_profile}
\end{figure}

We observe that $V^+$ is increasing with a constant rate for small
$y^+$ and then follows the log-law of the MDR.  The results for
$\lambda=1$ are shown in Fig. \ref{v_profile}.
We note that for $\nu_p=100$ the velocity profile first follows
the viscous profile and then joins the MDR until it begins to
cross back to the Newtonian plug. For $\nu_p$=1000 the viscous
layer is increased in size, but then the velocity profile becomes
parallel to the MDR but with a smaller intercept, signaling less
efficicient drag reduction. Once $\nu_p$ reaches the value of 10
000, the viscous layer becomes rather large, of the order of 20.
The velocity profile sill succeeds to run parallel the MDR, but with a
much reduced intercept.

\begin{figure}
  \includegraphics[width=3.5in]{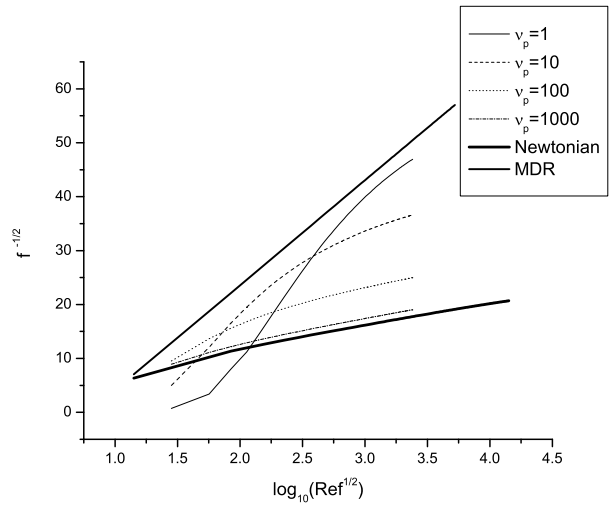}\\
  \caption{$f^{-1/2}$ as the function of $\log_{10}({\rm Re} f^{1/2})$ with
$\lambda=1$ with various values of
$\nu_p$.}\label{Redep}
\end{figure}

To compare the prediction of the model to the experimental results
shown in Fig. \ref{Virk} we need to relate  Re$_\tau$ to Re as
used in that figure. The relation is
\begin{equation}
\label{Re} {\rm Re}_L  \equiv Re_{\tau} V^+_L  \ ,
\end{equation}
where $V^+_L$ is the velocity at the center of the channel.  Fig.
\ref{Redep} shows $f^{-1/2}$ as a function of $\log_{10}({\rm
Re}_L f^{1/2})$ with various values of $\nu_p$. One can see that for large $\nu_p$, we obtain a drag enhancement when Re is
small.  This enhancement decreases with increasing Re and
eventually for large enough Re there is drag reduction. The
maximum amount of drag reduction, however, cannot exceed the MDR.
For small $\nu_p$, the drag enhancement is nearly unnoticeable.
However, drag reduction in this case is smaller for large Re
because the Newtonian plug occurred earlier. It should be noted that
the curves in Fig. \ref{Redep} exhibit the ``ladder"
characteristics for different values of $\nu_p$. This was identified
as ``Type B" drag reduction by Virk \cite{ASME}, and is the
fundamental characteristic of drag reduction by rod-like
polymers.  We see that our results agree reasonably well with the
experiments presented in Prandtl-K\'arm\'an coordinates.

\begin{figure}
  \includegraphics[width=3.5in]{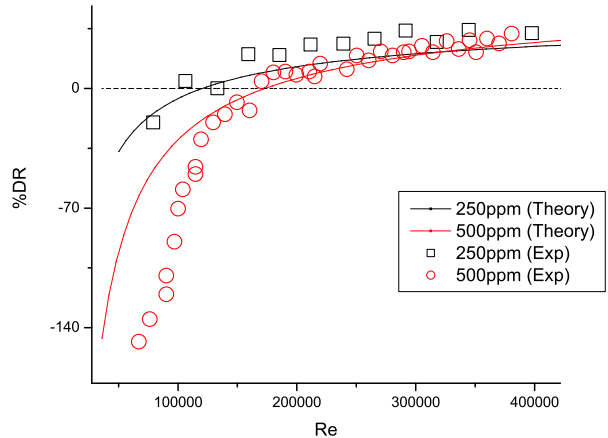}\\
  \caption{$f^{-1/2}$ as the function of $\log_{10}({\rm Re} f^{1/2})$ with
$\lambda=1$ with various values of $\nu_p$.}\label{DR}
\end{figure}

To compare the theory to the experimental results in DR(\%) vs Re coordinates
 we use the experimentally measured
values of $g$ and $\nu_p$ to predict the amount of drag reduction.
As we mentioned before, the effective viscosity in laminar flow of
XG solutions is well-approximated by Ostwald-de Waele model. This
implies in the low-Re flow, $g$ is given by (in MKS units):
\begin{equation}
g=k_1 S^{n-1} = k_1 \left(\frac{\nu_0 {\rm
Re}_\tau^2}{L^2}\right)^{n-1} \ , \label{eqg}
\end{equation}
where the last equality is obtained by multiplying  Eq.
({\ref{De}}) by $\gamma$.  The experimental value of $\nu_p$ for XG solution is
given by (in MKS units):
\begin{equation}
\nu_p= 0.011147 ~C^{1.422}\nu_0 \ , \label{eqnu}
\end{equation}
where $C$ is the concentration of XG in wppm.  The value of $f$ can then be calculated by: (i) obtaining
the experimental values of $C$, $k_1$ and $n$ for Eqs. (\ref{eqg})
and (\ref{eqnu}), (ii) rescaling the variables $S$, $K$ and $W$ by
Eq(\ref{rescale}) and finally, (iii) solving Eqs. (\ref{Fnueff}) and
(\ref{ceqn}) to get $f$ as a function of ${\rm Re}_L$. It should
be noted that in the present experiments the definition of Reynolds number is
in general different from our definition in (\ref{Re}). For the sake of comparison
of the theory with the experiments we assume that two Reynolds numbers are proportional to each other:
\begin{equation}
{\rm Re}_L = 12 {\rm Re}_{\rm exp} \ .
\end{equation}
Fig. {\ref{DR}} shows the comparison of the percentage of drag
reduction (enhancement) between the theoretical predictions and
the experimental results. The two data sets shown pertain to  $c=250$wppm
($k_1=11.04, \ n=0.727 $) and $c=500$wppm ($k_1=11.04, \ n=0.727
$). The agreement between theory and experiment is very satisfactory.

\section{Discussion and Conclusions}

In summary, we studied the Re dependence of the drag reduction by
rod-like polymers.  We showed that when the value of Re is small,
the drag is {\em enhanced} due to the homogenous increase of the
effective viscosity.  When Re is sufficiently large, due to the
turbulent activity, the effective viscosity varies as a function
of the distance to the wall.  As the result, the amount of drag is
reduced.  We use a simple interpolation between low Re
and the high Re flows to account for this. The resulting theoretical results agree semi-quantitatively
with the experiments both in Prandtl-K\'arm\'an coordinates and in
DR(\%) vs Re coordinates.

It should be noted that only rod-like polymers exhibit drag
enhancement for low Re.  For flexible polymers, the drag is
the same as that of the Newtonian flow for low Re, and only after a critical
value of Re, the drag reduction sets in \cite{Virk, Bonn2}.  This
difference in behavior from rod-like polymers
is because the flexible polymers are coiled when Re is small. They
do not affect the flow unless Re increases enough to allow the shear
to develop to affect the coil-stretch transition in the flexible polymers.
In contrast, rod-like polymers are always
extended and therefore they can affect the flow for all
Re.

\acknowledgments This work has been supported in part by the
European Commission under a TMR grant and by the US-Israel bi-national Science Foundation.

\end{document}